# Peculiarities of crystal structure of terbium (III) trifluoroacetate trihydrate

## Vyacheslav I. Belyi, Vladimir N.Ikorskii, Alexander A.Rastorguev, Angelina A.Remova

A.V.Nikolaev Institute of Inorganic Chemistry of Russian Academy of Sciences

3, Acad. Lavrentiev prospect, 630090 Novosibirsk, Russia

E-mail: a_rast@che.nsk.su

A comparative study of the crystalline and molecular structures of trifluoroacetate trihydrates of Tb(III) prepared from $Tb(OH)_3$ and $Tb_2(CO_3)_3$ by reaction with trifluoroacetic acid is made. Elemental analysis of the two reaction products in indicated them to be compounds of the same $C_6H_6F_9O_9Tb$. According to the X-ray structural analysis the molecular structure of the first product is described by the average statistical formula $Tb_2(CF_3COO)_5(CF_3COOH)(H_2O)_5(OH)$ (Tb-I) and the compound is a dimer with a centre of inversion. In the second case the product is also a centrosymmetric dimer but its formula is $Tb_2(CF_3COO)_6(H_2O)_6$ (Tb-II). The compounds only difference is in the crystallographic positions of their hydrogen atoms. A luminescent analysis using $Tb^{3+}$ $^5D_4 \rightarrow ^7F_0$ transition indicated the presence within the structure of Tb-I of the structure of Tb-II. Magnetic measurements showed that Tb-II has an antiferromagnetic ordering at low temperatures. In the case of Tb-I there also appears a ferromagnetic ordering. On the basis of the available experimental results such a behaviour was explained to be the result of some peculiar features of the hydrogen bonds in the molecules.

Introduction

The rare earth carboxylates exhibit a wide range of crystalline structures thanks to the large coordination numbers of the ions, the variety of the types of the carboxylate ligand coordination in complexes and the diversity of the hydrogen bonds appearing in individual compounds and their solid solutions [1-3]. The structure of these compounds can be monomeric, dimeric and polymeric. The rare earth carboxylates find application as starting materials in the synthesis of superconductors, magnetic materials and catalysts[3]. The luminescent properties of these compounds may be useful in the creation of luminescent materials[4]. The compounds of Eu(III) and Tb(III) are often used as probes due to the simple structure of their electronic transitions from excited $^5D_0$ and $^5D_4$ states, respectively, and the relatively long lifetimes of these states[5].

At the present time, the crystal structure of trifluoroacetate hydrates is known for almost all rare earth elements[6]. All these compounds have the same molecular structure of centrosymmetric dimers with four carboxylate ligands and is described by the formula $Ln_2(CF_3COO)_6(H_2O)_6$ except for the compounds of lanthanum and cerium which have a polymeric structure[6]. In this study we do not discuss "unexpected" products, as were observed in the reaction of $TbCl_3 \cdot 6H_2O$ with $[Na(hfa)]_n$, $hfa=CF_3COCHCOCF_3$ [2].

At the present time there exist two main synthetic routes leading to the rare earth trifluoroacetate trihydrates. The simplest one is to dissolve a rare earth oxide in an aqueous solution of trifluoroacetic acid and to evaporate the solution at room temperature until crystals are precipitated. In the second route $Ln(OH)_3$ is used instead of the oxide. Until recently both routes were believed to lead to the same result: the formation of products described by the formula $Ln(CF_3COO)_3(H_2O)_3$[7-8]. In this work the trifluoroacetate trihydrates of Tb(III) were prepared both from $Tb(OH)_3$ (Tb-I) and $Tb_2(CO_3)_3$ (Tb-II) and the properties of the resulting compounds were studied and compared with each other.

## EXPERIMENTAL

**$Tb_2(CF_3COO)_4(CF_3COOH)_2(H_2O)_4(OH)_2$ : synthesis**

Terbium hydroxide was prepared by the precipitation with ammonium hydroxide from an aqueous solution of a pure grade $TbCl_3 \cdot 6H_2O$. The crystals were filtered off and washed with a 1% aqueous solution of ammonia until the reaction for the presence of $Cl^-$ was negative. For the synthesis trifluoroacetic acid was added by drops to an aqueous suspension of the hydroxide until the suspension was completely dissolved and then an additional amount of the acid was added to obtain pH=1. The resulting solution was evaporated at room temperature until crystals were precipitated. The crystals were separated from the mother liquor, re-crystallised from water and dried to a constant mass. The compound derived in this way was designated as Tb-I, its elemental composition corresponded to the formula $C_6H_6F_9O_9Tb$[8].

**$Tb_2(CF_3COO)_6(H_2O)_6$ : synthesis**

The crystals of $Tb_2(CF_3COO)_6(H_2O)_6$ (further referred to as Tb-II) were synthesised as follows, To an aqueous solution of a pure grade $TbCl_3 \cdot$the equivalent amount of a reagent purity grade $NaHCO_3$ was added in small portions under constant stirring. The resulting finely crystalline precipitate of $Tb_2(CO_3)_3$ was filtered off, washed with a small quantity of water and dissolved in trifluoroacetic acid which was taken in a slight excess (to pH~2). The solution was evaporated at room temperature until crystals of the compound were formed. The crystals were separated from the mother liquor and dried in air to a constant weight. The elemental composition (C, H, F, O, Tb) of the compounds corresponded to the formula $C_6H_6F_9O_9Tb$[8].

**Luminescent analysis**

The luminescence spectra were measured on an open architecture automated apparatus that was used in the Ref.[9-11]. The details of the construction are given in the Ref.[14]. A mercury DRSH-250 lamp and a hydrogen LDD-400 lamp served as the sources of the exiting light. The exciting light was filtered with a MDR-23 monochromator. The luminescence was separated with a modified DFS-24 spectrometer and recorded with a FEU-100 photomultiplier at a spectral slit width of 0.07nm. The spectrum was scanned in steps of 0.001 nm. For the calibration with respect to the wavelengths a low-pressure mercury-helium lamp and the Balmier lines of hydrogen of a LDD-400 lamp were used. The measurements were done at room temperature. The mathematical processing of the spectra was performed with an exclusive program of Sergei V. Koshcheev[10]. The same finely crystallised powder samples were used in the luminescent, $^1H$ NMR and magnetic measurements.

**Magnetic measurements**

The magnetic properties of the Tb-I, Tb-II, Gd-1 and Gd-2 complexes were measured with a Quantum Design SQUID magnetometer. The measurements were performed in a magnetic field of 500 Oe at temperatures increasing from 2 to 300K. The effective magnetic moment was calculated by the formula $\mu_{eff} = ((3k/N_A\beta^2)\chi T^2)^{1/2} \approx (8\chi T)^{1/2}$, where $\chi$ is the molar magnetic susceptibility, $k$ is the Boltzman constant, $N_A$ is the Avogadro number, $\beta$ is the Bohr magneton.

## RESULTS AND DISCUSSION

The compounds of Tb(III) were synthesised by both procedures and their structural, spectral-luminescent and magnetic properties were studied for the first time[9-11]. In the case of terbium the first procedures proved to be problematic since terbium oxide always contains a quantity of a tetravalent terbium in the form of $TbO_2$. For this reason, $Tb_2(CO_3)_3$ was used as the starting compound. The colourless crystals were re-crystallised and the most suitable ones were selected for the X-ray diffraction study.

Table 1 shows the parameters of their hydrogen bonds for Tb-I and Tb-II. It was shown that both compounds have a monoclinic unit cell and the (space group P2$_1$/c) with very close dimensions of the unit cells (the parameters and the volumes of the two compounds differ in the second figure), suggesting them to be individual compounds with very close structures[11,12]. From an analysis of the electron density peak positions in the studied compounds it was possible to locate the hydrogen atoms of the water in the structure of the molecule. It was possible to locate the hydrogen atoms at the oxygen atoms of water O(1w) and O(2w). It was in Tb-1 that peculiar features in the electronic density positions were observed. The H(3wb) atom at 0(3w) and the H(31) at 0(31) were refined with a weight of ½ (Fig. 1). According to X-ray single crystal data the average statistical composition of Tb-I is described by the formula Tb$_2$(CF$_3$COO)$_5$(CF$_3$COOH)(H$_2$O)$_5$(OH). The dimer has a centre of inversion since the environments of its terbium ions were found to be crystallographically equivalent.

Fig. 1 shows the structure of an "average statistical" molecule. The intro- and intermolecular hydrogen bonds are indicated and the fluorine atoms omitted. The fluorides of all CF$_3$ occupy two statistically equiprobable positions in the both cases.

The hydrogen atoms in the Tb-II single crystal are reliably located with a statistical weight of 1. There are no peak corresponding to hydrogen atom near O(31) (the co-ordinated oxygen of the monodentate ligand) (Fig. 2). The composition of the compound is described by the formula Tb$_2$(CF$_3$COO)$_6$(H$_2$O)$_6$. The compound is isostructural with compounds of other rare earth elements. According to the X-ray analysis data the structures with Eu(III) and Gd(III), both the oxide- and hydroxide-based ones, do not differ and are described by the formula Ln$_2$(CF$_3$COO)$_6$(H$_2$O)$_6$. This shows that it is the compound of terbium prepared be the reaction of Tb(OH)$_3$ with trifluoroacetic acid that is unique.

The presence in Tb-I of the OH$^-$ groups bound to Tb$^{3+}$ is confirmed by the $^1$H NMR data. The number of the OH$^-$ hydrogens amounts to less than 10% relative to the content of all other hydrogens in the molecule, is in good agreement with the 1:11 ratio derived from the average statistical formula.

In our earlier study of the luminescence in Tb-I[9] the emission was shown to be mainly due the electronic transitions in the Tb$^{3+}$ ion from the lowest excited level $^5D_4$ to the $^7F_J$ levels. The luminescent spectra contain all transitions with J=6-0. From an analysis of the luminescence spectra, luminescence excitation spectra and absorption spectra it was possible to estimate the energies of the singlet and triplet levels of the molecule, their positions in the term structure of the Tb$^{3+}$ ion and the magnitudes of the nepheloxetic shift. There was lack of coincidence between the excitation spectra for the green ($\lambda \cong 544$ nm) and blue ($\lambda \cong 489$ nm) luminescence bands. The details of the terbium trifluoroacetate structure were obtained with the help of the $^5D_4 \to {}^7F_0$ transition. Although according to the X-ray structural data all Tb$^{3+}$ ions have the same crystallographic environments the number of the Stark components in this transition exceeded the expected value of 9. In Fig. 3 the luminescence spectra in the $^5D_4 \to {}^7F_0$ transition are compared for Tb-I (Fig. Curve a,c) and Tb-II (Fig. Curve b). The spectrum of Tb-I has two peaks, $\lambda_1 \cong 680$ nm and $\lambda_2 \cong 683$ nm, shifted by 60.7 cm$^{-1}$ relative to each other. The luminescence spectrum for Tb-II has only one peak at $\lambda_1 \cong 680$ nm. All three bands have the same half-width $\Delta\nu \cong 53$ cm$^{-1}$. The maximum $\lambda_1$ for Tb-I coincides with $\lambda_3$ for Tb-II. From this we conclude that the splitting is not caused by the interaction of the Tb$^{3+}$ ions in the dimer but is rather due to the different nearest environments of Tb$^{3+}$. The band with the maximum at $\lambda_1 \cong 680$ nm correspond to an environment of three main CF$_3$COO$^-$ ligands while the band at $\lambda_2 \cong 683$ nm is observed for the environment consisting of one OH$^-$ group and two main ligands. Thus, Tb-I is a combination of two complexes Tb(CF$_3$COO)$_2$(CF$_3$COOH)(H$_2$O)$_2$(OH) and Tb(CF$_3$COO)$_3$(H$_2$O)$_3$.

Fig. 4 shows the temperature behaviour of the effective magnetic moment $\mu_{\text{eff}}(T)$ and reverse magnetic susceptibility $\chi^{-1}(T)$ for Tb-I (Fig. Curve 1) and Tb-II (Fig. Curve 2). At high temperatures the value of $\mu_{eff}$ coincides with the value for the trivalent Tb ion. A noticeable feature in the Tb-I behaviour consists in the fact that $\mu_{eff}$ increases with decreasing temperature and begins to fall as the temperature decreases below 30 K. For Tb-II $\mu_{eff}$ decreases monotonically with temperature. For Tb-I the behaviour of the magnetic moment is describable by the Curie-Weiss equation $\chi(T)=C/(T-\theta)$ for two regions of temperatures: below and above 30 K (the inset in Fig. 4). For Tb-I the behaviour of $\chi(T)$ is described by one and the same equation over the whole temperature range of measurements from 2 to 300 K. The optimum parameters for the Curie-Weiss equation for both compounds are shown in Table 2. We made an attempt to relate the different magnetic behaviours of the complexes to some peculiarities of their structure which could lead to a change of the magnetic interaction. For this purpose, the energy parameters of the exchange interactions were estimated quantitatively by the molecular field method. In this method the Weiss constant $\theta$ is related to the exchange energy parameters $J_i$ by the relation $\theta = 2/3\ s(s+1) \sum z_i J_i$, where $z_i$ is the number of the first, second etc nearest neighbours, $s$ is the spin of the ion[13]. The values of $zJ$ (obtained taking into account only the nearest neighbour interactions) are shown in Table .2. For Tb-I two values of the exchange parameter differing in the sign and magnitude were obtained. The increase of $\mu_{eff}$ for Tb-I at temperatures from 30 to 300 K may be connected with the action of the ferromagnetic interaction which is stronger in this range of temperatures. With the further decrease in temperature it is the aniferromagnetic interaction that begins to be more effective making $\mu_{eff}$ decrease as the temperature falls below 30 K. In Tb-II the interaction in the dimer is entirely antiferromagnetic and has a magnitude comparable with that of the interaction of the second type inTb-I (see Table.2). For comparison, we have studied trifluoroacetate hydrates of gadolinium, $Gd_2(CF_3COO)_6(H_2O)_6$, which were prepared from the oxide (Curve 1, Fig. 5) and hydroxide (Curve 2, Fig.5). Both compounds are identical in structure; their magnetic characteristics are very close and indicate the presence of a week antiferromagnetic interaction. From the obtained data we can conclude that the ferromagnetic interactions observed in Tb-I are due to the additional hydrogen bonds that appear in this compound (Table.2).

## Conclusion

It is shown that the hydroxide-based trifluoroacetate trihydrate of Tb(III) is a combination of two complexes: $Tb(CF_3COO)_2(CF_3COOH)(H_2O)_2(OH)$ and $Tb(CF_3COO)_3(H_2O)_3$ in the ratio 1:1. The structural, magnetic and luminescent properties of this compound are different from those of $Tb_2(CF_3COO)_6(H_2O)_6$ obtained by the chemical reaction of $Tb_2(CO_3)_3$ with $CF_3COOH$. This allows us to consider it to be an individual compound. It should be noted that the unit cell parameters in Tb-I are smaller than in Tb-II[11]. These differences in the structure may be due to additional interactions in the dimer. In Tb-I, for example, it may be a dynamic disordering of the proton positions. The dynamic character of the additional hydrogen bonds in the structure of Tb-I is supported by the temperature behaviour of the magnetic susceptibility described above and by the change of the luminescence in the $^5D_4 \rightarrow {}^7F_0$ transition with decreasing temperature observed in[11] but the reason which leads to the formation of Tb-I is not yet clear. For other rare earth elements (Eu, Gd) no compounds of the type of Tb-I were found. And no dimer of the composition $Tb_2(CF_3COO)_4(CF_3COOH)_2(H_2O)_4(OH)_2$ could be isolated experimentally over a wide range of pH values. This means that the reason for the appearance of crystals of Tb-I lies in peculiar features of the kinetics of their self-assembling. It may be suggested that even at pH=1 the formation of crystals occurs from a solution containing both $Tb(CF_3COO)_3$ and not fully neutralised monomers $Tb(CF_3COO)_2(OH)$ with one $OH^-$ group. The fact that in the solution the molecules are present in the monomer rather than in the dimer form has been was demonstrated experimentally[8]. And it is the $Tb^{3+}$ ions which for some reason find it favourable for themselves to co-ordinate these monomers together with the molecules of water to produce a crystal with the structure and properties of Tb-I. Another explanation of the additional ferromagnetic interaction in Tb-I and the presence of two centres of luminescence (the two bands for $^5D_4 \rightarrow {}^7F_0$) may be the deviation from the stoichiometry with respect to hydrogen observed for $Tb_2(CF_3COO)_5(CF_3COOH)(H_x)(H_2O)_5(OH)$ (x=0,...,2)[12]. Indeed if the structural fragment of the composition $Tb(CF_3COO)_2(CF_3COOH)(H)(OH)(H_2O)_2$ actually exists the formation of $H_2$ and the additional hydrogen bonds could get stabilised by the scheme shown in Figure 6. The proof of this hypothesis requires further investigations.

## Acknowledgement.

The authors wish to thank Svetlana G. Kozlova, Doctor of Science (Phys.-Math) for the measurements of the $^1$H NMR spectra, Galina V. Romanenko Doctor of Science (Chem.) for the X-Ray diffraction measurements and useful discussions and Nina P. Sokolova, PhD, senior research worker for the synthesis of the compounds. The work was supported by a grant of the President of the Russian Federation Grant No НШ-1042.2003.3.

Table 1. Hydrogen bond parameters in Tb-I and и Tb-II structures (Å and degrees)[1]

| D-H...A | d(D-H) | d(H...A) | d(D...A) | <(DHA) |
|---|---|---|---|---|
| **Tb-I** | | | | |
| O(31)-H(31)...O(3W) | 0.70(7) | 2.26(8) | 2.836(1) | 141(10) |
| O(31)-H(31)...O(3W)#2 | 0.70(7) | 2.81(8) | 3.308(1) | 130(8) |
| O(1W)-H(1WA)...O(32)#3 | 0.59(3) | 2.18(1) | 2.766(2) | 169(1) |
| O(1W)-H(1WB)...F(22)#4 | 0.89(3) | 2.33(3) | 3.207(2) | 165(3) |
| O(2W)-H(2WA)...O(32) | 0.77(2) | 1.92(2) | 2.642(2) | 157(2) |
| O(2W)-H(2WA)...O(31) | 0.77(2) | 2.58(2) | 2.885(1) | 105(1) |
| O(2W)-H(2WB)...F(13)#1 | 0.67(2) | 2.56(2) | 3.144(3) | 146(2) |
| O(3W)-H(3WA)...O(11)#2 | 0.78(2) | 2.08(2) | 2.838(1) | 162(2) |
| O(3W)-H(3WA)...F(11)#2 | 0.78(2) | 2.53(2) | 3.079(2) | 129(2) |
| O(3W)-H(3WA)...F(11A)#2 | 0.78(2) | 2.53(2) | 3.067(2) | 128(2) |

| Tb-II | | | | |
|---|---|---|---|---|
| O(1W)-H(1WA)...O(32)#2 | 0.58(8) | 2.18(8) | 2.764(7) | 175(10) |
| O(1W)-H(1WB)...F(22)#3 | 0.95(14) | 2.29(14) | 3.232(13) | 175(11) |
| O(2W)-H(2WA)...O(32) | 0.69(10) | 2.05(11) | 2.651(6) | 146(11) |
| O(2W)-H(2WA)...O(31) | 0.69(10) | 2.69(10) | 2.888(6) | 99(9) |
| O(2W)-H(2WB)...F(13)#1 | 0.83(12) | 2.51(11) | 3.148(18) | 135(10) |
| O(3W)-H(3WA)...O(11)#4 | 0.76(11) | 2.11(10) | 2.853(5) | 163(11) |
| O(3W)-H(3WA)...F(11)#4 | 0.76(11) | 2.57(11) | 3.10(3) | 128(9) |
| O(3W)-H(3WA)...F(11A)#4 | 0.76(11) | 2.58(11) | 3.08(3) | 125(9) |
| O(3W)-H(3WB)...F(32A)#4 | 0.68(8) | 2.36(8) | 2.982(14) | 154(8) |

Table 2. Optimum parameters of the Curie-Weiss equation.

| | | C (cm$^3$ K/mole) | θ (K) | μ | zJ (cm$^{-1}$) |
|---|---|---|---|---|---|
| 1 | 2-300 K | 10.76±0.02 | -1.48±0.2 | 9.28 | -0.13 |
| 2 | <30 K | 13.0±0.02 | -1.34±0.02 | 10.2 | -0.12 |
| | >30 K | 10.92±0.02 | 6.17±0.4 | 9.35 | 0.54 |

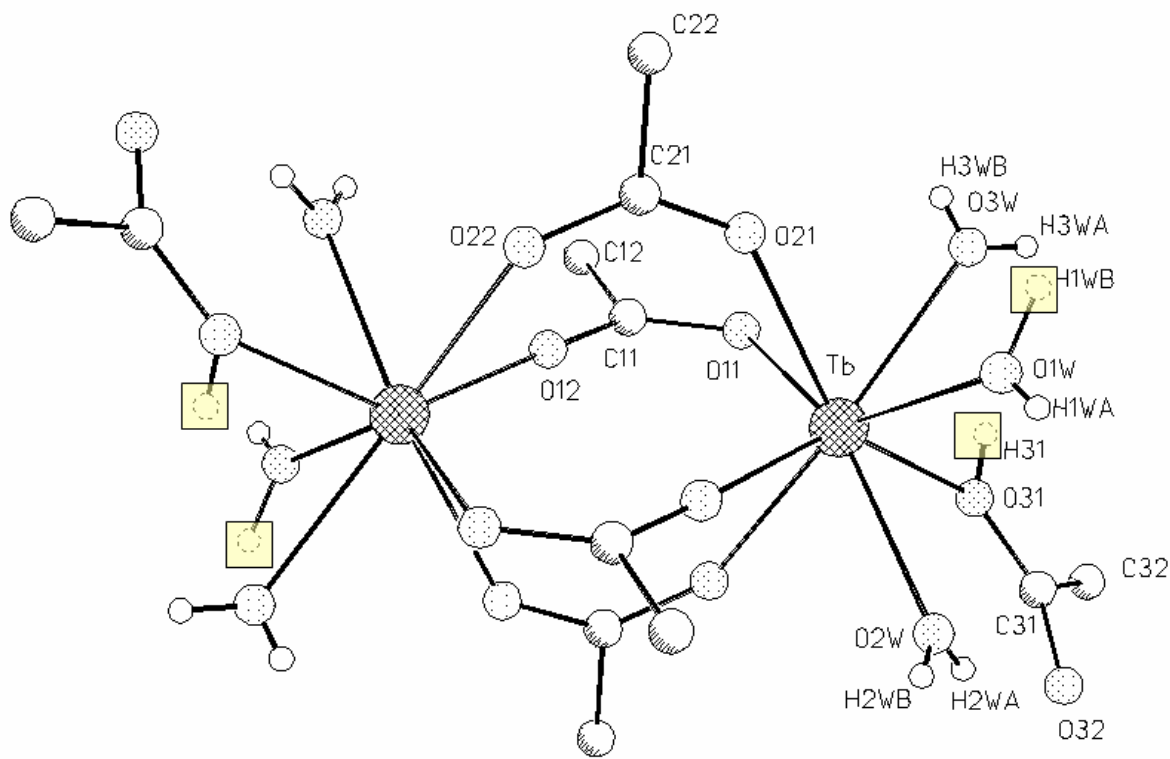

Fig.1. Structure of molecule Tb-I synthesised from terbium(III) hydroxide. The squares depict hydrogen atoms with structure occupation factor ½.. All fluorine atoms have been omitted for clarity.

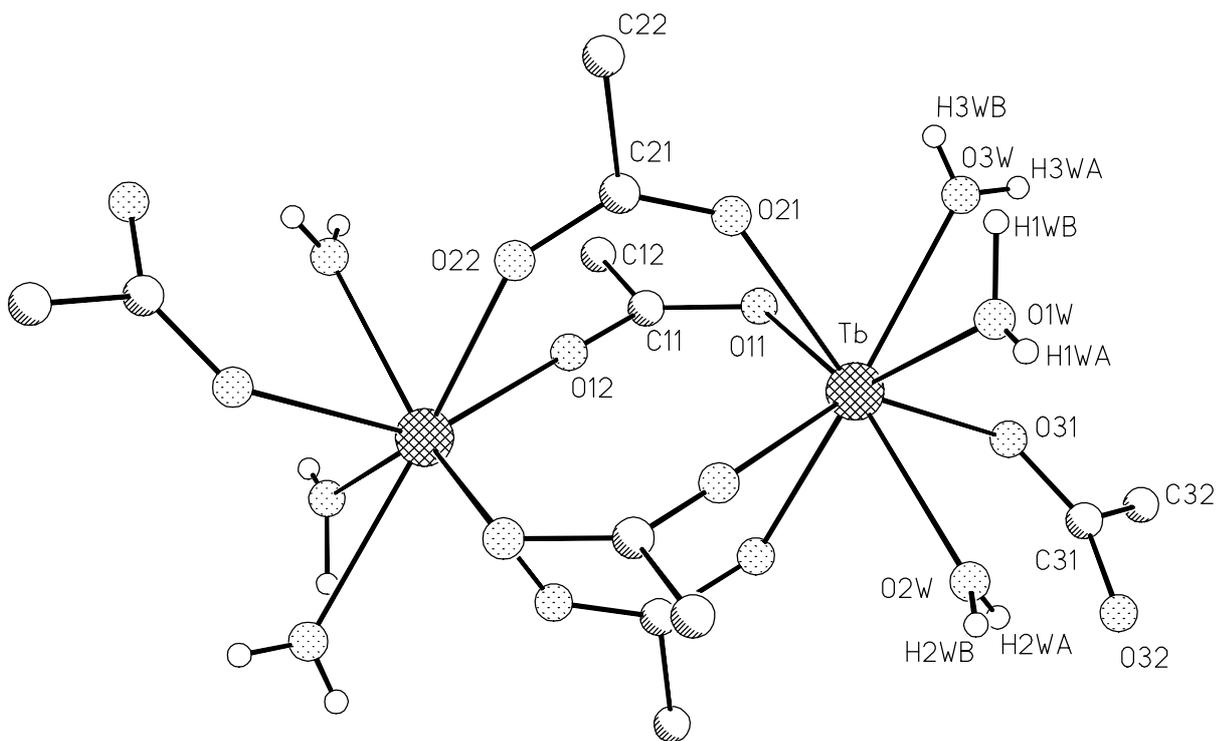

Fig.2. Molecule $Tb_2(CF_3COO)_6(H_2O)_6$ of the terbium carbonate-derived compound (Tb-II) according to the X-ray structural analysis. All fluorine atoms have been omitted for clarity.

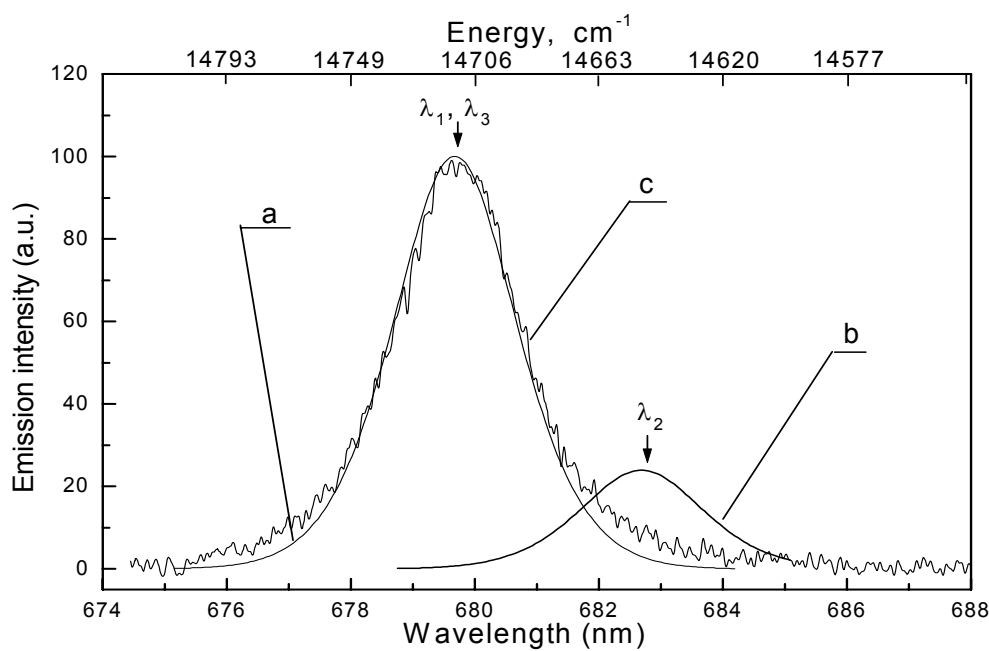

Fig.3. Luminescence spectra of the $^5D_4 \to {}^7F_0$ transition in the ion $Tb^{3+}$ ($\lambda_{exc.} \cong 366$ nm). The curves **a**, **c** are the Gauss-Lorenz decomposition components of the curve [10] for the Tb-I. Compound. Curve **b** is an experimental curve for the Tb-II compound.

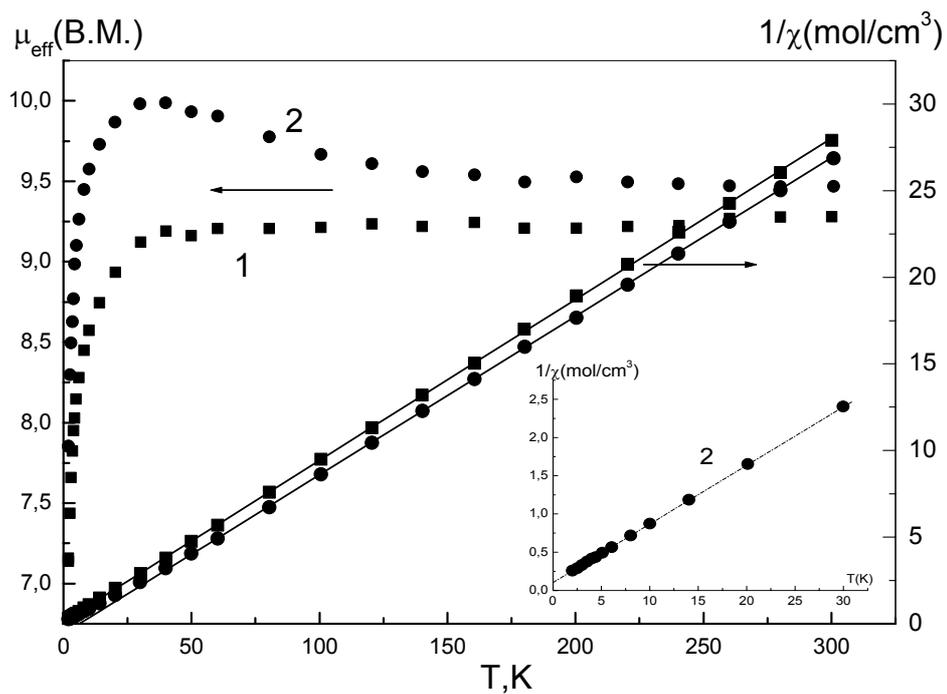

Fig.4. The $1/\chi(T)$ and $\mu_{eff}(T)$ dependencies for the Tb-I(2) and Tb-II(1) complexes. Shown in the inset is the $1/\chi(T)$ dependence in the temperature range from 2 to 30 K for Tb-I(2).

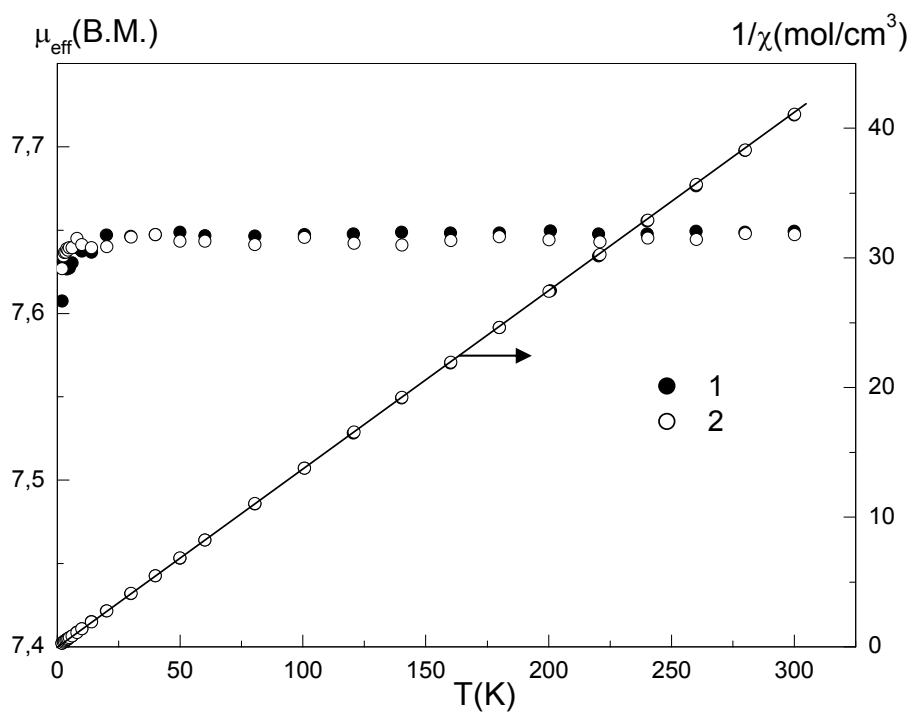

Fig.5. The $1/\chi(T)$ and $\mu_{eff}(T)$ dependencies for the complexes of Gd.

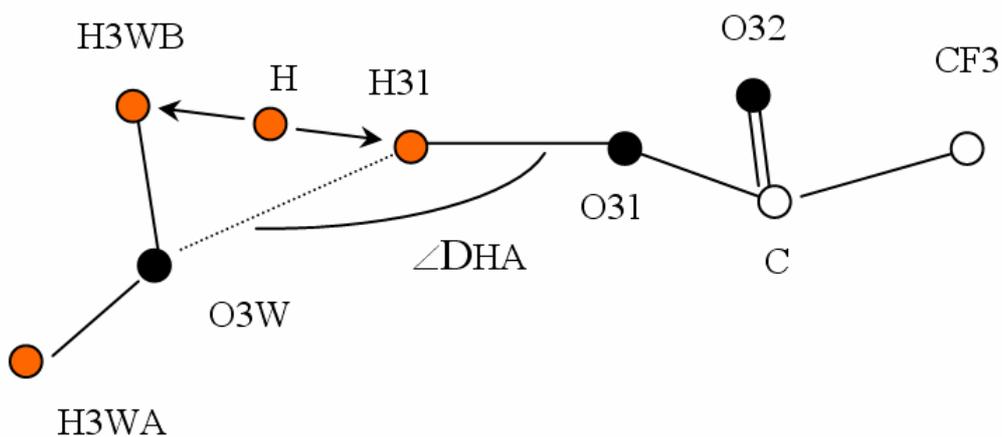

Fig.6. The structure of the intramolecular O(31)-H(31)...O(3W)#2 hydrogen bond (see Table .1) and an "extra" migrating hydrogen H.

---

[1] The symmetry operation codes for the generation of the equivalent atoms: #1 -x+1,-y+2,-z+1; #2 -x+2,-y+2,-z+1; #3 x,-y+3/2,z+1/2, #4 -x+2,-y+2,-z+2